\documentclass{optica-article}

\journal{opticajournal} 

\articletype{Research Article}

\begin{document}

\title{Field Test of Quantum Key Distribution with High Key Creation Efficiency}

\author{Yung-Cheng Kao,\authormark{1,4,$\dagger$} Sheng-Hsuan Huang,\authormark{1,4} ,Chin-Hsuan Chang,\authormark{1}Chih-Hsiang Wu,\authormark{1}Shih-Hsien Chu,\authormark{1}Jian Jiang,\authormark{1}An-Chi Zhang,\authormark{1} Sheng-Yao Huang,\authormark{1} Jhih-Heng Yan,\authormark{2}Kai-Ming Feng,\authormark{2,3} and Chih-Sung Chuu\authormark{1,3,*}}

\address{\authormark{1}Department of Physics, National Tsing Hua University, Hsinchu 30013, Taiwan\\
\authormark{2}Institute of Communications Engineering, National Tsing Hua University, Hsinchu 30013, Taiwan\\
\authormark{3}Center for Quantum Technology, Hsinchu 30013, Taiwan\\
\authormark{4}Thess authors contributed equally to this work}

\email{\authormark{$\dagger$}yungchengkao@gapp.nthu.edu.tw}
\email{\authormark{*}cschuu@nthu.phys.edu.tw} 


\begin{abstract*} 
Quantum key distribution (QKD) promises unconditional security for communication. However, the random choices of the measurement basis in QKD usually result in low key creation efficiency. This drawback is overcome in the differential-phase-shift QKD, provided that each photon can be prepared in a large number of time bins with a proper waveform. In this work we develop a miniature 1550-nm single-photon source to generate narrowband single photon in 50 time bins with a nearly optimal waveform for achieving unity key creation efficiency. By utilizing these single photons in the field test, we demonstrate the differential-phase-shift QKD with a key creation efficiency of 97\%. Our work shows that the practical QKD can benefit from the narrowband single photons with controllable waveforms.

\end{abstract*}

\section{Introduction}
Quantum key distribution (QKD) promises unconditionally secure communication granted by the laws of physics \cite{Gisin02}. The first demonstration of QKD \cite{Ben92}, proposed by Bennett and Brassard \cite{Ben84}, was based on the polarization encoding. Since then, different protocols \cite{Bennett92,Inoue02,Stucki09,Lo2012,Sasaki2014,Boaron2018,Lucamarini2018} have been demonstrated to avoid the polarization distortion in optical fibers and extend the transmission distance \cite{Marand1995,Xiao05,Hiroki07,Peng2007,Xai09,Peev2009,Liu2010,Stucki2011,Sasaki2011,Wang2012,Korzh2015,Yin2016,Liao2017,Yin17,Islam2017,Boaron18,Bunandar2018,Minder2019,Chen2021}. While the weak coherent pulses (WCP) were widely used in these demonstrations due to their simple realization, the quantum light sources offer additional benefits. For example, the use of single photons can tolerate higher loss \cite{Lu00,Wang08} and higher quantum bit error rate (QBER) \cite{Wang2015}. The use of entangled photons can extend the distance of QKD with the help of quantum repeaters~\cite{Briegel98}. Recently, Liu \textit{et. al} \cite{Liu13} showed that the key creation efficiency (KCE) can be enhanced with single photons of long coherence time. By using an 795-nm heralded single-photon source based on the four-wave mixing in cold atoms \cite{Du08}, a KCE of 66.6\% was achieved in the laboratory environment with unconditional security.

To realize the long-distance QKD in fiber transmission system, photons in the telecommunication band are necessary for minimizing the photon loss in the fibers. In this paper we demonstrate a miniature 1550-nm single-photon source based on the monolithic doubly resonant parametric down-conversion \cite{Chuu11}. The long coherence time and narrow bandwidth of the single photons not only increase the KCE but also reduce the QBER~\cite{Hon11}. In addition, we show that the double-exponential waveform of the single photons plays a key role to enhance the KCE. By shaping each photon into 50 equally spaced time bins, we demonstrate the field test of the differential-phase-shift (DPS) QKD with 97\% KCE and a QBER below the threshold level of unconditional security~\cite{Wen09} using an inter-university optical fiber network. Our work shows that the single photons with long coherence time and controlled waveforms are feasible for practical QKD.
\section{High key creation efficiency}

\begin{figure}[t]
\includegraphics[width=\linewidth]{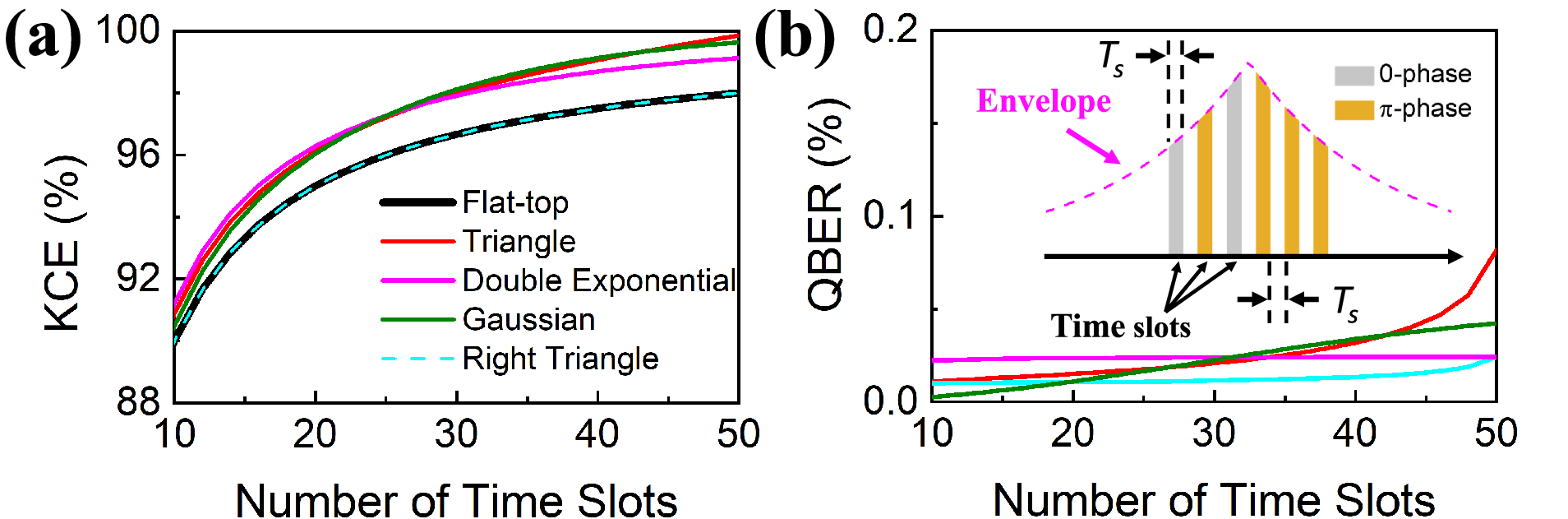}
\caption{\label{fig:1}(a) The key creation efficiencies and (b) the QBERs introduced by the non-flat waveforms for different numbers of time bins and waveforms of single-photon wavepackets.}
\end{figure}

In the DPS QKD \cite{Inoue02}, Alice (the sender) prepares each photon in multiple time bins. The bits (0 or 1) are then randomly encoded in the phase difference (0 or $\pi$) between the adjacent time bins. After receiving the photons, Bob (the receiver) measures the bits with an \textit{one-bit} delay Mach-Zehnder interferometer. If the photon is not detected in the first or last time slots, the bits are extracted by correlating the exit port of the photon to the phase difference. The KCE (namely, the probability of successfully creating one bit per photon) is thus equal to the probability of finding the photon in the time slots other than the first and last slots after the interference. In Fig.~\ref{fig:1}(a) we calculate the KCE for different numbers of time bins $N$ and waveforms. While high number of time bins is important to obtain high KCE (which approaches 100\% at very large $N$), the waveforms have a notable impact on the KCE. In general, the symmetrically decaying waveforms such as the triangle, double exponential, and Gaussian waveforms increase the KCE by lowering the probability of finding the single photons in the first and last time slots. However, due to the mismatch in the amplitudes between the adjacent time bins, the interference visibility of all non-flat-top waveforms deviate from 100\%. Nevertheless, the QBERs introduced by the amplitude mismatch are negligible compared to the threshold level 4.12\% of the unconditionally secure DPS QKD \cite{Wen09}. This can be seen in Fig.~\ref{fig:1}(b), where the QBER or the probability that a photon exits the wrong ports due to the imperfect visibility is calculated. 

\section{Miniature telecom single-photon source with controllable waveforms}

\begin{figure}[t!]
\includegraphics[width=0.95\linewidth]{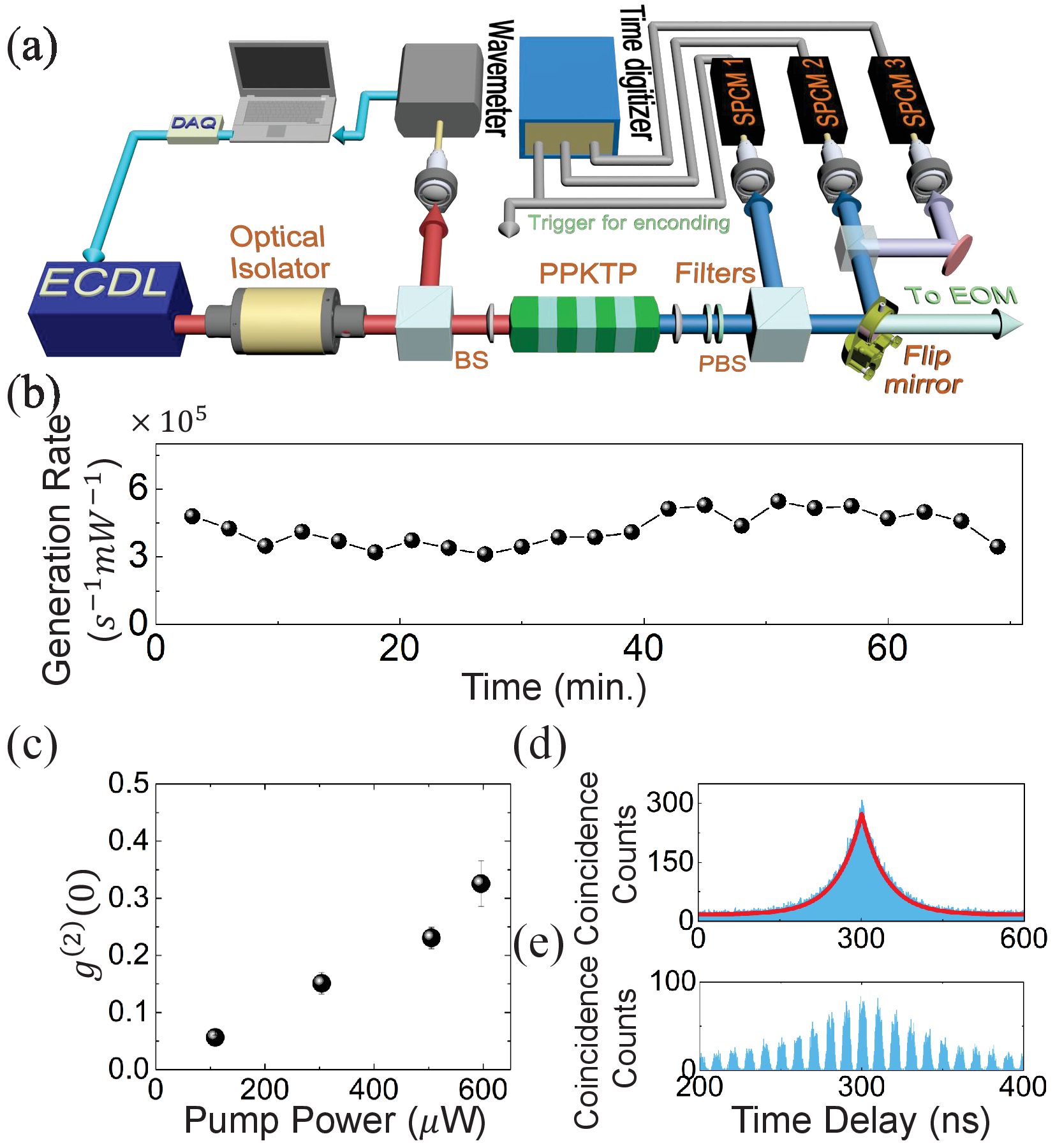}
\caption{\label{fig:2}(a) Experimental setup of the miniature 1550-nm single-photon source based on the monolithic doubly resonant parametric down-conversion. The abbreviations stand for the data acquisition (DAQ), beamsplitter (BS), polarizing beamsplitter (PBS), and electro-optic modulator (EOM). (b) The generation rate versus the time. (c) The pump power is kept low to ensure good single-photon quality ($g^{(2)}(0) < 0.5$). Due to the finite dead time ($\sim 2\ \mu$s) and the lack of the photon-number resolvability of the detectors, the measured $g^{(2)}(0)$ saturates when the pump power is above 600 $\mu$W. (d) The temporal wavepacket (blue area) of the heralded single photons. The red curve is the fit of the double-exponential function. (e) The electro-optic modulation of the single-photon wavepacket into a pulse train. The integration times are 180 s and 120 s in (d) and (e), respectively.}
\end{figure}

To increase the KCE, temporally long and symmetrically decaying single-photon wavepackets are advantageous. Fig.~\ref{fig:2}(a) illustrates the setup of our single-photon source. Degenerate photon pairs at 1550 nm are generated from a periodically poled and type-II phase-matched KTiOPO$_4$ (PPKTP) crystal inside a monolithic resonator, which is implemented by spherically polishing the end faces of the crystal and depositing a 1550-nm high-reflection coating. The pump is a 775-nm cw external cavity diode laser (ECDL) frequency-stabilized by a high-precision wavemeter (1-MHz resolution and 30-MHz absolute accuracy). By depositing an additional high-reflection coating at the pump wavelength on the rear end, we realize the double-pass pumping. Together with the monolithic resonator, this enables the photon pairs to be generated in a single longitudinal mode without the need of external filters \cite{Chuu11,Chuu12,Wu17}. After the photon pair are separated by a polarizing beamsplitter (PBS), a single (signal) photon is heralded by detecting an idler photon with a single-photon counting module (SPCM, ID Quantique ID230). Accounted for the fiber-coupling efficiency (75\%), transmission loss (5\%), and detector's quantum efficiency (15\%), the generation rate is $4.2 \times 10^5$~s$^{-1}$ per mW of the pump power with a fluctuation within $15 \%$ about the average rate [Fig.~\ref{fig:2}(b)]. Fig.~\ref{fig:2}(c) shows the measured second-order quantum coherence function at zero time delay $g^{(2)}(0)$ \cite{Gra86} of the heralded single photons. For the pump power below 600~$\mu$W, the antibunching ($g^{(2)}(0) < 0.5$) is clearly evident.

As shown in Fig.~\ref{fig:2}(d) (blue area), the wavepacket of the single photons has a long temporal length and a double-exponential waveform. The wavepacket is described by the Glauber correlation function \cite{Chuu11},
\begin{equation}
 G^{(2)}(\tau)=R^2+\frac{4\kappa ^2\Gamma_s \Gamma_i}{(\Gamma_s+\Gamma_i)^2}\times \left \{\begin{array}{ll}
                 e^{\Gamma_s\tau}, & \mbox{if $\tau<0$} \\ 
                 e^{-\Gamma_i\tau}, & \mbox{if $\tau>0$}
                \end{array} \right.
\label{eq:G2}
\end{equation}
where $\tau$ is the time delay between the detection of the signal and idler photons, $\kappa$ is the parametric coupling constant, and $R$ is the generation rate. By fitting the measured wavepacket to $G^{(2)}(\tau)$ [red curve in Fig.~\ref{fig:2}(d)], the $1/e^2$ width of the wavepacket is found to be 198.8 ns with the total cavity decay rates of the signal and idler photons being $\Gamma_s=1/(45.3\ {\rm ns})$ and $\Gamma_i=1/(44.1\ {\rm ns})$, respectively. To implement the time bins, the heralded single photons are temporally delayed by an optical fiber and then shaped by a fiber-coupled electro-optic intensity modulator using a square wave. An example of the modulated wavepacket is shown in Fig.~\ref{fig:2}(e), where the number of time bins can be controlled by the prescribed modulation function. The optical delay is necessary for synchronizing the arrival of the single photons at the modulator and the electro-optic modulation, which is triggered by the detection of the idler photons. The long temporal length also infers a nannorwband bandwidth \cite{Chuu11} of $\Delta f=[(\sqrt{\Gamma^4_s+6\Gamma^2_s\Gamma^2_i+\Gamma^4_i}-\Gamma^2_s-\Gamma^2_i)/2]^{1/2}/2\pi= 2.3$~MHz and a spectral brightness of $1.8 \times 10^5$~s$^{-1}$mW$^{-1}$MHz$^{-1}$. The narrow bandwidth is beneficial for reducing the QBER of the DPS QKD because the probability distribution of a photon's frequency causes the photon to exit at the wrong port of the one-bit delay Mach-Zehnder interferometer, which acts as a filter in the frequency domain with a sinusoidal transmittance. For our single photons, the QBER induced by the narrow bandwidth \cite{Hon11} is estimated to be $\int_{-\infty}^{\infty} \sin^2(\pi x/f_{\rm FSR}) \Delta f /\{2\pi [x^2 + (\Delta f/2)^2]\} dx \cong 1.2\%$, where $f_{\rm FSR} = 0.3$~GHz is the free spectral range of the Mach-Zehnder interferometer.

\section{Field test of high-KCE DPS QKD}

\begin{figure}[t]
\includegraphics[width=1\linewidth]{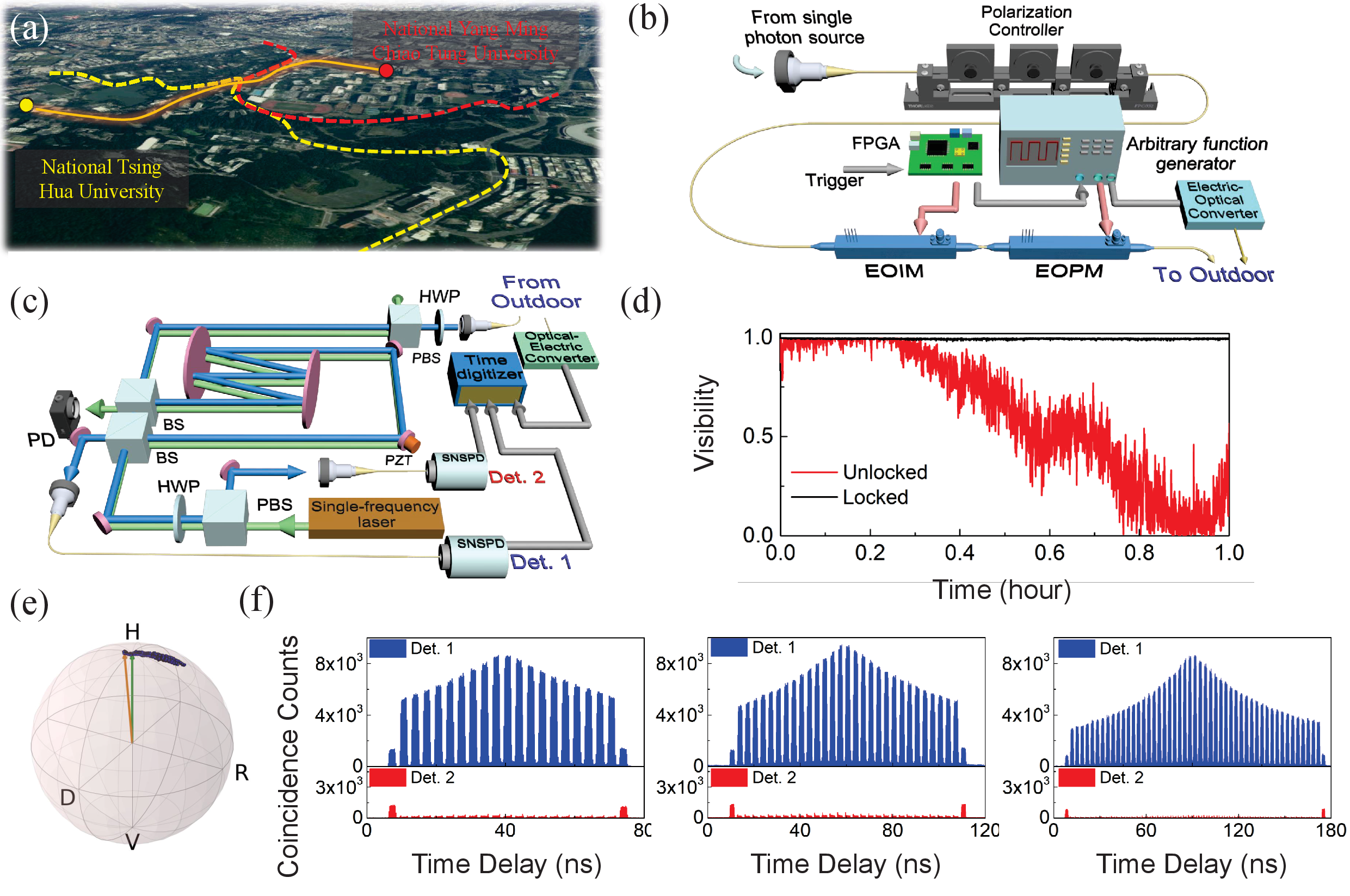}
\caption{\label{fig:3}(a) The field test exploits the inter-university optical fiber network (orange line) between the General II Building (yellow dot) at NTHU (bounded by the yellow dash line) and the IT Service Center (red dot) at NYCU (bounded by the red dash line) in Hsinchu City, Taiwan. Alice's and Bob's setup are illustrated in (b) and (c), respectively. The abbreviations stand for half-wave plate (HWP), beamsplitter (BS), polarizing beamsplitter (PBS), piezoelectric actuator (PZT), photodiode (PD), and superconducting nanowire single-photon detector (SNSPD). (d) The interference visibility of the Mach-Zehnder interferometer with (black curve) and without (red curve) the active stabilization. (e) The Poincar\'e sphere shows the fluctuation of a single-frequency laser's polarization state after passing through the fiber pair. The polarization state starts at the green arrow (the horizontal) and ends up in the orange arrow after 24 hours. (f) The time-resolved interference of single photons prepared in 20, 30, and 50 time bins (from the left to right). The integration time and coincidence time bin are 600 s and 0.2 ns, respectively. }
\end{figure}

The field test of the DPS QKD is carried out in the daytime using a pair of the inter-university fibers between the campuses of National Tsing Hua University (NTHU) and National Yang Ming Chiao Tung University (NYCU) in the Hsinchu City, Taiwan. The fiber pair, both extending from the General II Building in NTHU to the IT Service Center in NYCU [Fig.~\ref{fig:3}(a)], are connected at the NYCU site to form a fiber loop with a total length of 3.4 km. The attenuation of the fiber pair is 2 dB mainly due to the existing fiber-to-fiber connectors in the network. Fig.~\ref{fig:3}(b) illustrates Alice's setup at the NTHU site. The single photons, after being shaped into multiple time bins (square waves of 1/(300 MHz) period) by a fiber-coupled electro-optic intensity modulator (EOIM) with an insertion loss of 1.7 dB, are immediately sent into a fiber-coupled electro-optic phase modulator (EOPM) with an insertion loss of 1.6 dB. Alice uses the phase modulator to code the bits by applying random phase differences 0 and $\pi$ on adjacent time bins. To account for the broadening of the time bins by the modulator's finite bandwidth, the bin width of the phase modulation is chosen to be wider than the bin width of the intensity modulation.

The bits are later decoded by Bob with an one-bit delay Mach-Zehnder interferometer [Fig.~\ref{fig:3}(c)] at the NTHU site. To keep the interferometer stable during the field test, the long arm (total length of 1 m) is folded to reduce the overall size. In addition, the interference of a counter-propagating single-frequency laser, offseted in position to avoid the light leaking into the detectors, is continuously monitored. The interference signal is then fed to a piezoelectric actuator on a mirror to actively stabilize the optical path difference. As shown in Fig.~\ref{fig:3}(d), the interference visibility with the active stabilization (black curve) is nearly unchanged compared to that without the stabilization (red curve). In practice, the fluctuation of polarization in the fibers can also degrade the interference visibility. Such fluctuatione is shown in Fig.~\ref{fig:3}(e), where a single-frequency laser's polarization state (initially prepared in the horizontal) after transmitting through the fiber loop is recorded over 24 hours by a polarimeter. To achieve high visibility and low QBER, a polarizing beamsplitter is placed in front of the interferometer as a polarization filter, resulting in a reduction of the count rate and key rate by 2\%.  

\begin{figure}[t!]
\includegraphics[width=\linewidth]{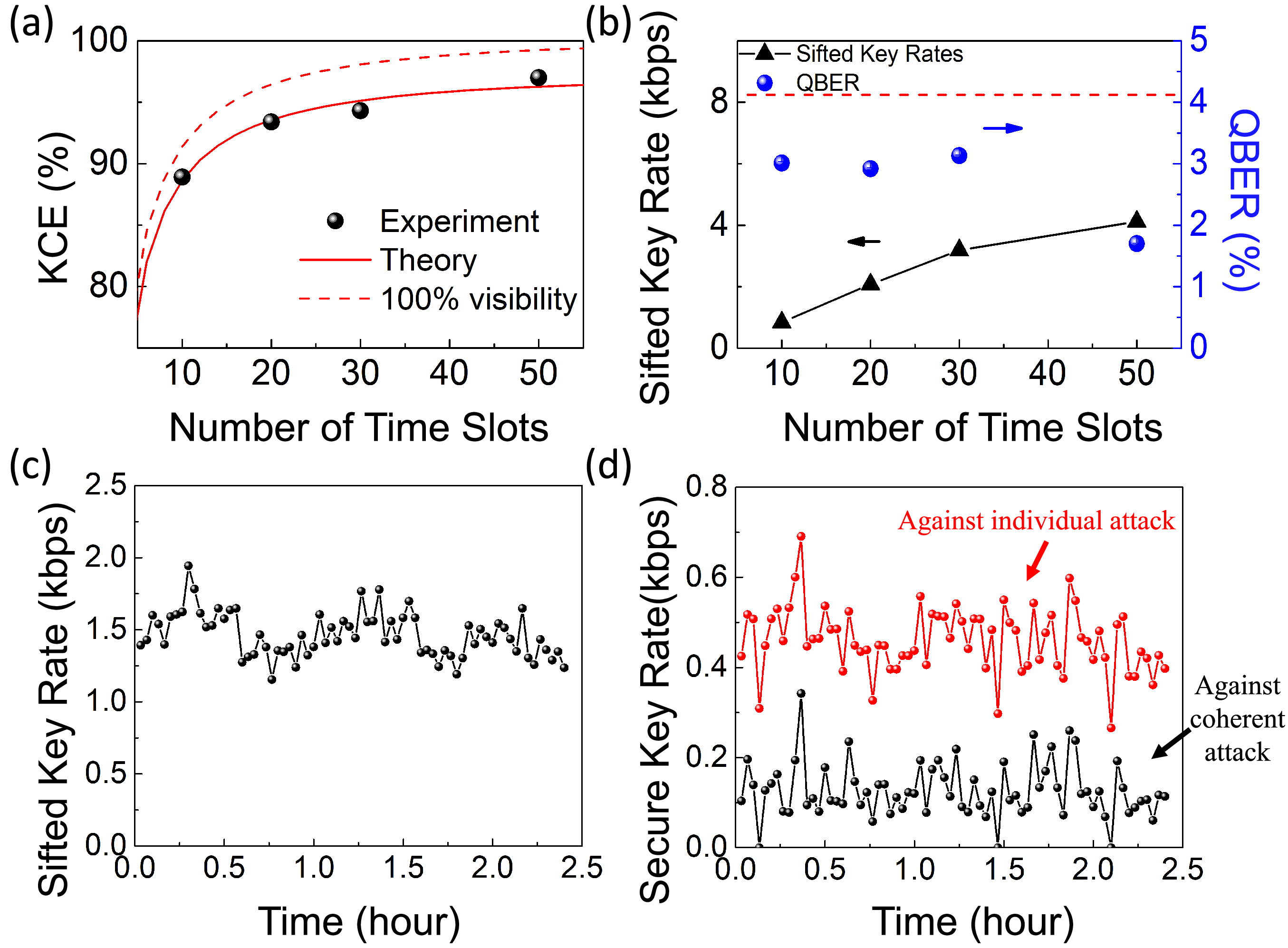}
\caption{\label{fig:4}(a) The key creation efficiency (dots) versus the number of time bins. The solid and dash curves correspond to the theory considering realistic and 100\% interference visibility, respectively. (b) The sifted key rates and QBERs versus the number of time bins. The dash line indicates the threshold level of unconditional security.}
\end{figure}

After a single photon enters the first beamsplitter of the interferometer, the wavefunctions taking the short and long paths interfere at the second beamsplitter. Depending on the phase difference between the adjacent time bins, constructive or destructive interference occurs between the time bins of the two wavefunctions, resulting in a click at one detector in the corresponding time slot. The time tags of these detection events are then sent back to Alice through a public channel for generating the bit string. Examples of the single-photon interference are shown in Fig.~\ref{fig:3}(f) for single photons prepared in 20, 30, and 50 time bins. All phase differences between the adjacent time bins are set to 0 in these examples. By analyzing these time-resolved interferences, some useful parameters can be obtained. Fig.~\ref{fig:4}(a) (dots) shows the KCE, which is given by the success probability of the interference, for different numbers of time bins. When the single photons are prepared in 50 time bins, the KCE reaches 97\% and is in good agreement with the theory (solid curve). The failure probability of the interference (excluding the first and last time slots), on the other hand, can cause the photons to show up at the wrong exit ports, thus introducing the QBER. Such QBER is shown in Fig.~\ref{fig:4}(b) (circular dots) for different numbers of time bins. Finally, as the photons taking part in the interference contribute to the key, the count rate of the single photons gives an estimate of the achievable sifted key rates, provided that the modulation bandwidths are not limited. Such sifted key rates are also plotted in Fig.~\ref{fig:4}(b) (triangles) for different number of time bins. The sifted key rate is higher at higher number of time bins because a larger portion of the single-photo wavepacket is used and the KCE is also higher.

In the demonstration of the field test, we use the field-programmable gate array (FPGA) and an arbitrary function generator to generate the time-bin and random phase modulation functions for Alice. The single-photon counting modules used in Bob's Mach-Zehnder interferometer and Alice's heralded single-photon source are also replaced by the superconducting nanowire single-photon detectors (Single Quantum Eos), which have a quantum efficiency higher than $82\%$ at 1550 nm. With 50 time bins and 97\% KCE, we obtain an average sifted key rate of $R_s = 3048$~bps and QBER of $e_b = 3.2\%$, which is below the threshold level of the unconditionally secure DPS QKD \cite{Wen09}. Assuming an error correction coding efficiency approaching the Shannon limit and an ideal single-photon source, the secure key rates are $R_s\{-\log_2[1-e_b^2-(1-6e_b)^2/2]-H(e_b)\}=1117$~bps against the general attack of individual photons \cite{Wen09,Waks06} and $R_s[1-H(e_b)-H((3+\sqrt{5})e_b)]=433$~bps against the coherent attack \cite{Wen09}, where $H(x)=-x\log_2x-(1-x)\log_2(1-x)$ is the binary Shannon entropy. To estimate the influence of the multiphoton emission, we approximate the photon statistics of our heralded single-photon source by a Poisson distribution (weak coherent state) with an average photon number of 0.37 \cite{Waks06}. We find that the secure key rate would decrease by 2\%.
	
\section{Conclusion}

We have shown that both the single-photon waveform and the number of time bins in DPS QKD play important roles in enhancing the KCE. To generate the single photons with the waveform for achieving the perfect KCE, we implement a miniature single-photon source based on the monolithic doubly resonant parametric down-conversion. The single-photon wavepacket has a nearly ideal waveform (double exponential) with a $1/e^2$ width of 198.8 ns and a generation rate of $4.2 \times 10^5$~s$^{-1}$mW$^{-1}$. With each photon shaped into 50 time bins, we demonstrate the field test of DPS QKD with 97\% KCE and a QBER below the threshold level of unconditional security. By increasing the number of time bins and interference visibility, which require faster modulation and shorter optical path difference of the interferometer, the KCE can be further increased [see the dash line in Fig.~\ref{fig:4}(b)] and the QBER can be reduced. Our work shows that the practical QKD can benefit from the single photons with long coherence time, narrow bandwidth, and controlled waveforms. Recently, a QKD network with a trusted relay have been set up across the campuses of NTHU and NYCU in Taiwan. The QKD system demonstrated here will be incorporated in this network to work with other QKD systems based on the weak coherent states.

\begin{backmatter}
\bmsection{Funding}
 This work was supported by the Center for Quantum Technology and the National Science and Technology Council, Taiwan (110-2112-M-007-021-MY3 and 111-2119-M-007-007).

\bmsection{Acknowledgments}
The authors would like to thank the Computer and Communication Center of NTHU and the Information Technology Service Center of NYCU for their kind assistance with the inter-university optical fiber network. We also thank C.-Y. Mou, H.-S. Goan, I. A. Yu, C.-M. Li, A. Yabushita, Y.-C. Liang, Y.-N. Chen, R.-K. Lee, and C.-M. Wu for the fruitful discussion and experimental assistance. 

\bmsection{Disclosures}
The authors declare no conflicts of interest.
\bmsection{Data availability}
Data underlying the results presented in this paper are not publicly available at this time but may be obtained from the authors upon reasonable request

\end{backmatter}


\providecommand{\noopsort}[1]{}\providecommand{\singleletter}[1]{#1}%






\end{document}